\documentclass[journal=jpccck,manuscript=article]{achemso}
\pdfoutput=1

\usepackage{graphicx}
\usepackage{dcolumn}
\usepackage{bm}
\usepackage{amsmath}
\DeclareMathOperator{\Tr}{Tr}

\author{Constantine Yannouleas}
\affiliation{School of Physics, Georgia Institute of Technology,
             Atlanta, Georgia 30332-0430}
\author{Igor Romanovsky}
\affiliation{School of Physics, Georgia Institute of Technology,
             Atlanta, Georgia 30332-0430}
\author{Uzi Landman}
\email{Uzi.Landman@physics.gatech.edu}
\affiliation{School of Physics, Georgia Institute of Technology,
             Atlanta, Georgia 30332-0430}

\title{Transport, Aharonov-Bohm, and Topological  Effects in Graphene Molecular Junctions and Graphene Nanorings}

\begin{document}

\begin{abstract}
The unique ultra-relativistic, massless, nature of electron states in two-dimensional extended graphene sheets,
brought about by the honeycomb lattice arrangement of carbon atoms in two-dimensions, provides ingress to
explorations of fundamental physical phenomena in graphene nanostructures. Here we explore the emergence of
new behavior of electrons in atomically precise segmented graphene nanoribbons (GNRs) and graphene rings with
the use of tight-binding calculations, non-equilibrium Green's function transport theory, and a newly
developed Dirac continuum model that absorbs the valence-to-conductance energy gaps as position-dependent
masses, including topological-in-origin mass-barriers at the contacts between segments. Through transport
investigations in variable-width segmented GNRs with armchair, zigzag, and mixed edge terminations we uncover
development of new Fabry-P\'{e}rot-like interference patterns in segmented GNRs, a crossover from  the
ultra-relativistic  massless regime, characteristic of extended graphene systems, to a massive relativistic
behavior in narrow armchair GNRs, and the emergence of nonrelativistic behavior in zigzag-terminated
GNRs. Evaluation of the electronic states in a polygonal graphene nanoring under the influence of an applied
magnetic field in the Aharonov-Bohm regime, and their analysis with the use of a relativistic quantum-field
theoretical model, unveils development of a topological-in-origin zero-energy soliton state and charge
fractionization. These results provide a unifying framework for analysis of electronic states, coherent
transport phenomena, and the interpretation of forthcoming experiments in segmented graphene nanoribbons and
polygonal rings.
\end{abstract}

\section{Introduction}

In the last three decades transport through molecular junctions \cite{nitz03,joac05,lind07,hod06,berg13,rai11}
has attracted much attention because of
fundamental aspects of the processes involved, as well as of potential practical prospects. In particular,
studies in this direction have intensified since the discovery of new forms of carbon allotropes, starting with
the fullerenes \cite{krot85} and carbon nanotubes \cite{iiji91} (CNTs) in the 1980s and 1990s
respectively, and the isolation of graphene \cite{geim04} in 2004. The above carbon allotropes differ in shape 
(curvature), topology, and dimensionality, with the fullerenes being zero-dimensional (0D) with spherical or 
prolate shape, the CNTs being one-dimensional (1D) cylinders, and graphene being a two-dimensional (2D) plane 
(or 1D planar ribbons \cite{fuji96,cohe06,wang07,waka10}). In the fullerenes the
carbon-atom network is made of (non-adjacent) hexagons and pentagons, whereas the CNTs and graphene are entirely
hexagonal lattices (curved in CNTs) described in terms of a unit cell with a two-atom basis with the two
carbon atoms occupying two sublattices ($A$ and $B$, also mapped into the up and down pseudospin states 
\cite{cast09}). In the absence of defects, in-plane ($\sigma$) bonding occurs through $sp^2$ hybrid orbitals
and out-of-plane bonding ($\pi$) involves the $p_z$ orbital; in the following only the physics of $\pi$-states
is considered. 

The hexagonal network topology of graphene gives rise to relativistic behavior of the low-energy excitations which 
is captured by the ultra-relativistic massless Dirac-Weyl (DW) equation, with the Fermi velocity of graphene 
($v_F = c /300$) replacing the velocity of light. \cite{cast09} 
Among the many manifestations of the relativistic behavior in
graphene is the Klein paradox, that is ``... unimpeded penetration of relativistic particles through high and
wide potential barriers -- is one of the most exotic and counterintuitive consequences of quantum
electrodynamics.''\cite{kats06} The surprising relativistic behavior in graphene has indeed been recognized in
the 2010 Nobel award in physics to A. Geim and K. Novosolev.  

Another carbon-based system that was the subject of an earlier (2000) chemistry Nobel award (to
A.J. Heeger, A.G. MacDiarmid, and H. Shirakawa) is conductive polymers, with polyacetelyne (PAC) being a
representative example. \cite{heeg88} PAC is a 1D chain of carbon atoms forming a conjugated polymer with $sp^2p_z$
hybridization that leads to one unpaired electron per carbon atom (half-filled $\pi$-band) as in
graphene. Linearizing the spectrum of this 1D equally-spaced carbon chain at the Fermi level (that is at the
Dirac points, i.e., the zero-energy points of the  energy vs momentum  dispersion relation) results in a 1D
massless Dirac-equation description of the low energy excitations of the system. \cite{jack81,jack12}

However, the equally-spaced 1D system is unstable and consequently it distorts (structurally) spontaneously
(Peierls distortion \cite{peie55}) yielding a modulation (alternation) of the spacing between successive
sites that results in dimerization of successive atoms along the chain and the opening of a gap in the
electronic spectrum. This dimerization can occur in two energetically degenerate, but spatially distinct,
patterns, termed as equivalent domain structures. Either of these domains is a realization of the 1D Dirac
equation with a constant mass term, ${\cal M}$. Disruption of the dimerization pattern (e.g., by  transforming 
from one domain to another along the chain) creates a domain wall which can be described in the 1D Dirac 
formulation through the use of a position-dependent mass term of alternating sign (with the spatial mass-profile
connecting $+{\cal M}$ with $-{\cal M}$). The solution of this generalized 1D Dirac equation is a soliton  
characterized by having zero energy and by being localized at the domain wall. In this paper, the Dirac equation 
with position-dependent mass will be used in investigations of the electronic structure, transport, and 
magnetic-field-induced phenomena (Aharonov-Bohm) in graphene nanostructures such as nanoribbon junctions and
rings. \cite{roma12,roma13,yann14}

From the above we conclude that in the two extreme size-domains, that is a 2D infinite graphene sheet and a
1D carbon chain (PAC), the systems exhibit behavior that is relativistic in nature. In an attempt to bridge
between these two size-domains, we briefly discuss in the following systems of successively larger width,
starting from the polyacene (a quasi-1D chain of fused benzene rings) which may be regarded as the narrowest
graphene nanoribbon (GNR) with zigzag edge terminations. 

Polyacene was investigated \cite{kive83} first in 1983. It was found that, as in graphene and PAC
(without distortion), the valence and conduction bands of undistorted polyacene touch at the edge of the
Brillouin zone. However, unlike 2D graphene and PAC, the dispersion relation about the touching point is
quadratic, conferring a non-relativistic (Schr\"{o}dinger equation) character. We show in this paper that this
surprising finding persists for sufficiently narrow GNRs with zigzag termination (zGNRs). However, narrow
armchair-terminated GNRs (aGNRs) are found here to maintain relativistic behavior, with metallic ones being massless
and semiconducting ones being massive (both classes obeying the Einstein energy relation).

It this paper, we discuss mainly manifestations of relativistic and/or nonrelativistic quantum behavior
explored through theoretical considerations of transport measurements in segmented graphene nanoribbons of 
variable width, and spectral and topological effects in graphene nanorings in the presence of magnetic fields.

Transport through narrow graphene channels $-$ particularly bottom-up fabricated and atomically-precise
graphene nanoribbons \cite{ruff10,fuhr10,huan12,derl13,nari14,hart14,sini14,ruff12} $-$ is expected to offer
ingress to unique behavior of Dirac electrons in graphene nanostructures. In particular,
the wave nature of elementary particles (e.g., electrons and photons) is commonly manifested and demonstrated
in transport processes. Because of an exceptionally high electron mobility and a long mean-free path, it has
been anticipated that graphene \cite{cast09} devices hold the promise for the realization, measurement, and
possible utilization of fundamental aspects of coherent and ballistic transport behavior, which to date have been
observed, with varying degrees of success, mainly at semiconductor interfaces \cite{hou89,ji03}, quantum point
contacts \cite{vwee88}, metallic wires \cite{pasc95}, and carbon nanotubes \cite{lian01}.

Another manifestation of coherent ballistic transport are interference phenomena, reflecting the
wave nature of the transporting physical object, and associated most often with optical (electromagnetic
waves, photons) systems or with analogies to such systems (that is, the behavior of massless particles, as
in graphene sheets). Measurements of interference patterns are commonly made with the use of interferometers,
most familiar among them the multi-pass optical Fabry-P\'{e}rot (OFP) interferometer \cite{vaugbook,lipsbook}.
The advent of 2D forms of carbon allotropes has motivated the study of optical-like interference
phenomena associated with relativistic massless electrons, as in the case of metallic carbon nanotubes
\cite{lian01} and graphene 2D $p$-$n$ junctions \cite{rick13}. (We note that the hallmark of the OFP is that
the energy separation between successive maxima of the interference pattern varies as the inverse of the
cavity length $L$.)

For GNRs with segments of different widths, our investigations reveal diverse Fabry-P\'{e}rot transport modes 
beyond the OFP case, with conductance quantization steps ($nG_0, n = 1, 2, 3, \ldots$, with $G_0=2e^2/h$)
found only for uniform GNRs. In particular, three distinct categories of Fabry-P\'{e}rot interference patterns 
are identified:
\begin{enumerate}
\item
{\it FP-A:\/} An {\it optical\/} FP pattern corresponding to {\it massless\/} graphene electrons
exhibiting {\it equal spacing\/} between neighboring peaks. This pattern is associated with
metallic armchair nanoribbon central segments. This category is subdivided further to {\it FP-A1\/} and
{\it FP-A2} depending on whether a valence-to-conduction gap is absent ({\it FP-A1}, associated with
metallic armchair leads), or present ({\it FP-A2}, corresponding to semiconducting armchair leads).
\item
{\it FP-B:} A {\it massive relativistic\/} FP pattern exhibiting a shift in the conduction onset due to the
valence-to-conduction gap and {\it unequal peak spacings\/}. This pattern is associated with semiconducting
armchair nanoribbon central segments, irrespective of whether the leads are metallic armchair, semiconducting
armchair, or zigzag.
\item
{\it FP-C:} A {\it massive non-relativistic\/} FP pattern with $1/L^2$ {\it peak spacings\/}, but with a
vanishing valence-to-conduction gap, $L$ being the length of the central segment. This pattern is the one
expected for usual semiconductors described by the (nonrelativistic) Schr\"{o}dinger equation, and it is 
associated with zigzag nanoribbon central segments, irrespective of whether zigzag or metallic armchair leads 
are used.
\end{enumerate}

We report in this paper on the unique apects of transport through segmented GNRs
obtained from tight-binding non-equilibrium Green's function \cite{dattabook,datt92} (TB-NEGF) calculations in
conjunction with an analysis based on a one-dimensional (1D) relativistic Dirac continuum model.
This continuum model goes beyond the physics of the massless Dirac-Weyl (DW) electron, familiar from
two-dimensional (2D) honeycomb carbon sheets \cite{cast09}, and
it is referred to by us as the Dirac-Fabry-P\'{e}rot (DFP) theory (see below for the choice of name). In
particular, it is shown that the valence-to-conduction energy gap in armchair GNR (aGNR) segments, as well as
the barriers at the interfaces between nanoribbon segments, can be incorporated in an effective position-dependent
mass term in the Dirac hamiltonian; the transport solutions associated with this hamiltonian exhibit conductance
patterns comparable to those obtained from the microscopic NEGF calculations. For zigzag graphene nanoribbon
(zGNR) segments, the valence-to-conduction energy gap vanishes, and the mass term is consonant with excitations
corresponding to massive nonrelativistic Schr\"{o}dinger-type carriers. The faithful reproduction of these
unique TB-NEGF conductance patterns by the DFP theory, including mixed armchair-zigzag configurations (where
the carriers transit from a relativistic to a nonrelativistic regime), provides a unifying framework for
analysis of coherent transport phenomena and for interpretation of experiments targeting fundamental
understanding of transport in GNRs and the future development of graphene nanoelectronics.

To demonstrate the aforementioned soliton formation due to structural topological effects (discussed by us
above in the context of polyacetylene), we explore with numerical tight-binding calculations and a 
Dirac-Kronig-Penney (DKP) approach, soliton formation and charge fractionization in graphene rhombic rings;
this DKP approach is based on a generalized Dirac equation with alternating-sign position-dependent masses. 

Before leaving the Introduction, we mention that, due to their importance as fundamental phenomena, 
Aharonov-Bohm-type effects in graphene-nanoribbons systems have attracted (in addition to Refs.\ 
\citenum{roma12,roma13,yann14}) considerable theoretical attention.
\cite{hod06,ezaw07,rech07,baha09,wurm10,nguy13,nguy14} These latter theoretical papers, however, based their 
analysis exclusively on tight-binding and/or DFT calculations\cite{hod06,ezaw07,baha09,nguy13,nguy14}, or they
used in addition a two-dimensional Dirac equation with infinite-mass boundary conditions.\cite{rech07,wurm10} 
Transcending the level of current understanding which explores direct similaritiess with the Aharonov-Bohm physics
in semiconductor and metallic mesoscopic rings, our work here analyzes the TB results in conjunction with a 
continuum 1D generalized Dirac equation (that incorporates a position-dependent mass term), and thus it enables 
investigations of until-now unexplored topological aspects and relativistic quantum-field analogies of the AB 
effect in graphene nanosystems.   

Furthermore we note that oscillations in the conductance of graphene nanoribbons in the presence of magnetic
barriers were found in a theoretical study \cite{xu08} (using exclusively a TB-NEGF approach), as well as
in an experimental investigation of high quality bilayer nanoribbons, \cite{jiao10} and they were attributed
to Fabry-P\'{e}rot-type interference. In the absence of a continuum Dirac analysis, the precise relation of such 
oscillatory patterns to our Fabry-P\'{e}rot theory (based on the incorporation of a position-dependent mass
term in the Dirac equation at zero magnetic field) warrants further investigation. 

Finally, gap engineering in graphene ribbons under strong external fields was studied in Ref.\ \citenum{ritt08}. We
stress again that one of the main results in this paper is the appearance of ``forbidden'' solitonic states 
inside the energy gap in the context of the low-magnetic-field Aharonov-Bohm spectra of graphene nanorings;
see the part titled ``Aharonov-Bohm spectra of rhombic graphene rings'' in the Results section.       

\section{Methods}

\subsection{Dirac-Fabry-P\'{e}rot model}
The energy of a fermion (with one-dimensional 
momentum $p_x$) is given by the Einstein relativistic relation 
$E=\sqrt{ (p_x v_F)^2+({\cal M} v_F^2)^2 }$, where ${\cal M}$ is the rest mass and $v_F$ is the
Fermi velocity of graphene. In a gapped uniform graphene nanoribbon, the mass parameter is related to the
particle-hole energy gap, $\Delta$, as ${\cal M}=\Delta/(2 v_F^2)$. Following the relativisitic 
quantum-field Lagrangian formalism, the mass ${\cal M}$ is replaced by a position-dependent scalar 
Higgs field $\phi(x) \equiv m(x)v_F^2$, to which the relativistic fermionic field $\Psi(x)$ couples 
through the Yukawa Lagrangian \cite{roma13} ${\cal L}_Y = -\phi \Psi^\dagger \beta \Psi$ 
($\beta$ being a Pauli matrix). For $\phi(x) \equiv \phi_0$ (constant) ${\cal M} v_F^2=\phi_0$, and 
the massive fermion Dirac theory is recovered. The Dirac equation is generalized as (here we do not
consider applied electric or magnetic fields)
\begin{equation}
[E-V(x)] \Psi + i \hbar v_F \alpha \frac{\partial \Psi}{\partial x} - \beta \phi(x) \Psi=0.
\label{direq}
\end{equation}
In one dimension, the fermion field is a two-component spinor $\Psi = (\psi_u, \psi_l)^T$;
$u$ and $l$ stand, respectively, for the upper and lower component and $\alpha$ and $\beta$ can be any 
two of the three Pauli matrices. Note that the Higgs field enters in the last term of Eq.\ (\ref{direq}).
$V(x)$ in the first term is the usual electrostatic potential, which is inoperative due to the Klein 
phenomenon and thus is set to zero for the case of the armchair nanoribbons (where the excitations 
are relativistic). The generalized Dirac Eq. (\ref{direq}) is used in the 
construction of the transfer matrices of the Dirac-Fabry-P\'{e}rot model described below.

The building block of the DFP model is a 2$\times$2 wave-function matrix ${\bf \Omega}$ formed by the
components of two independent spinor solutions (at a point $x$) of the onedimensional, first-order
generalized Dirac equation [see Eq.\ (3) in the main paper]. 
${\bf \Omega}$ plays \cite{mcke87} the role of the Wronskian matrix ${\bf W}$ used in the second-order 
nonrelativistic Kronig-Penney model. Following Ref.\ \citenum{mcke87} and generalizing to $N$
regions, we use the simple form of ${\bf \Omega}$ in the Dirac representation
($\alpha=\sigma_1$, $\beta=\sigma_3$), namely
\begin{equation}
{\bf \Omega}_K (x) = \left( \begin{array}{cc}
e^{i K x} & e^{-i K x} \\
\Lambda e^{i K x}& -\Lambda e^{-i K x} \end{array} \right),
\label{ome}
\end{equation}
where
\begin{equation}
K^2=\frac{(E-V)^2-m^2 v_F^4}{\hbar^2 v_F^2}, \;\;\; \Lambda= \frac{\hbar v_F K}{E-V+m v_F^2}.
\label{klam}
\end{equation}
The transfer matrix for a given region (extending between two matching points $x_1$ and $x_2$
specifying the potential steps $m_i^{(n)}$) is the product
${\bf M}_K (x_1,x_2)= {\bf \Omega}_K (x_2){\bf \Omega}_K^{-1} (x_1)$;  this latter matrix depends
only on the width $x_2-x_1$ of the region, and not separately on $x_1$ or $x_2$.

The transfer matrix corresponding to a series of $N$ regions can be formed \cite{roma13} 
as the product
\begin{equation}
{\bf t}_{1,N+1} = \prod_{i=1,N} {\bf M}_{K_i} (x_i,x_{i+1}),
\label{tside}
\end{equation}
where $|x_{i+1}-x_i|=L_i$ is the width of the $i$th region [with $(m,V,K,\Lambda) \rightarrow
(m_i,V_i,K_i,\Lambda_i)$]. The transfer matrix associated with the
transmission of a free fermion of mass ${\cal M}$ (incoming from the right) through the 
multiple mass barriers is the product
\begin{equation}
{\bf {\cal T}}(E) = {\bf \Omega}^{-1}_k(x_{N+1}) {\bf t}_{1,N+1} {\bf \Omega}_k(x_1),
\label{thex}
\end{equation}
with $k=\sqrt{(E-V)^2-{\cal M}^2 v_F^4}/(\hbar v_F)$, $|E-V| \geq {\cal M} v_F^2$; for armchair leads
$V=0$, while for zigzag leads $V=\mp {\cal M} v_F^2$. Naturally, in the case of metallic armchair leads, 
$k=E/(\hbar v_F)$, since ${\cal M}=0$.

Then the transmission coefficient $T$ is
\begin{equation}
T=\frac{1}{|{\cal T}_{22}|^2},
\label{trans}
\end{equation}
while the reflection coefficient is given by
\begin{equation}
R=\left| \frac{{\cal T}_{12}}{{\cal T}_{22}}\right| ^2.
\label{refl}
\end{equation}

At zero temperature, the conductance is given by $G = (2e^2/h) T$;
$T$ is the transmission coefficient in Eq.\ (\ref{trans}).\\

\subsection{Dirac-Kronig-Penney superlatice model}

The transfer matrix corresponding to either half of the rhombus can be formed \cite{roma13} 
as the product
\begin{equation}
{\bf t}_n = \prod_{i=1,3} {\bf M}_K (x_i,x_{i+1}),\;\;\; x_1=0,\; x_4=L,
\label{tside2}
\end{equation}
with $L$ being the length of half of the perimeter of the rhombus; $L=L_1+L_2+L_3$, with $L_1=L_3=a$ and $L_2=b$.
The transfer matrix associated with the complete unit cell (encircling the rhombic ring) is the product
\begin{equation}
{\bf T}=\prod_{n=1}^2 {\bf t}_n.
\label{thex2}
\end{equation}

Following  Refs.\ \citenum{roma13,imry83}, we consider the superlattice generated from the virtual
periodic translation of the unit cell as a result of the application of a magnetic field $B$
perpendicular to the ring. Then the Aharonov-Bohm energy spectra are given as solutions of the
dispersion relation
\begin{equation}
\cos \left[ 2\pi(\Phi/\Phi_0+\eta) \right] = \Tr[{\bf T}(E)]/2,
\label{disrel}
\end{equation}
where we have explicitly denoted the dependence of the r.h.s. on the energy $E$; $\eta=0$ for a rhombus with
type-I corners and $\eta=1/2$ for a rhombus with type-II corners.

The energy spectra and single-particle densities do not depend on a specific representation.
However, the wave functions (upper and lower spinor components of the fermionic field $\Psi$) do
depend on the representation used. To transform the initial DKP wave functions to the
($\alpha=\sigma_2$, $\beta=\sigma_1$) representation, which corresponds to the natural separation of
the tight-binding amplitudes into the $A$ and $B$ sublattices, we apply successively the unitary
transformations $D_{23}=(\sigma_2+\sigma_3)/\sqrt{2}$ and $D_3=\exp(i \pi \sigma_3/4)$.

\subsection{TB-NEGF formalism}

To describe the properties of graphene nanostructures in the tight-binding approximation, we use the 
hamiltonian 
\begin{equation}
H_{\text{TB}}= - t \sum_{<i,j>} c^\dagger_i c_j + h.c.,
\label{htb}
\end{equation}
with $< >$ indicating summation over the nearest-neighbor sites $i,j$. $t=2.7$ eV
is the hopping parameter of two-dimensional graphene.

To calculate the TB-NEGF transmission coefficients, the Hamiltonian (\ref{htb}) is employed in conjunction 
with the well known transport formalism which is based on the nonequilibrium Green's functions
\cite{dattabook,datt92}.

According to the Landauer theory, the linear conductance is $G(E)=(2 e^2/h) T(E)$,
where the transmission coefficient is calculated as $T(E)=\Tr[\Gamma_L {\cal G} \Gamma_R {\cal G}^\dagger]$.
The Green's function ${\cal G}(E)$ is given by 
\begin{equation}
{\cal G}(E) = (E+i \eta - H_{\text{TB}}^{\text{dev}} - \Sigma_L - \Sigma_R)^{-1},
\label{ttb}
\end{equation}
with $H_{\text{TB}}^{\text{dev}}$ being the Hamiltonian of the isolated device (junction without the
leads). The self-energies $\Sigma_{L(R)}$ are given by 
$\Sigma_{L(R)} = \tau_{L(R)}[E+i \eta - H_{\text{TB}}^{L(R)}]^{-1}\tau_{L(R)}^\dagger$,
where the hopping matrices $\tau_{L(R)}$ describe the left (right) device-to-lead coupling,
and $H_{\text{TB}}^{L(R)}$ is the Hamiltonian of the semi-infinite left (right) lead. 
The broadening matrices are given by $\Gamma_{L(R)}=i[\Sigma_{L(R)}-\Sigma_{L(R)}^\dagger]$.   

\begin{figure*}
\centering\includegraphics[width=14cm]{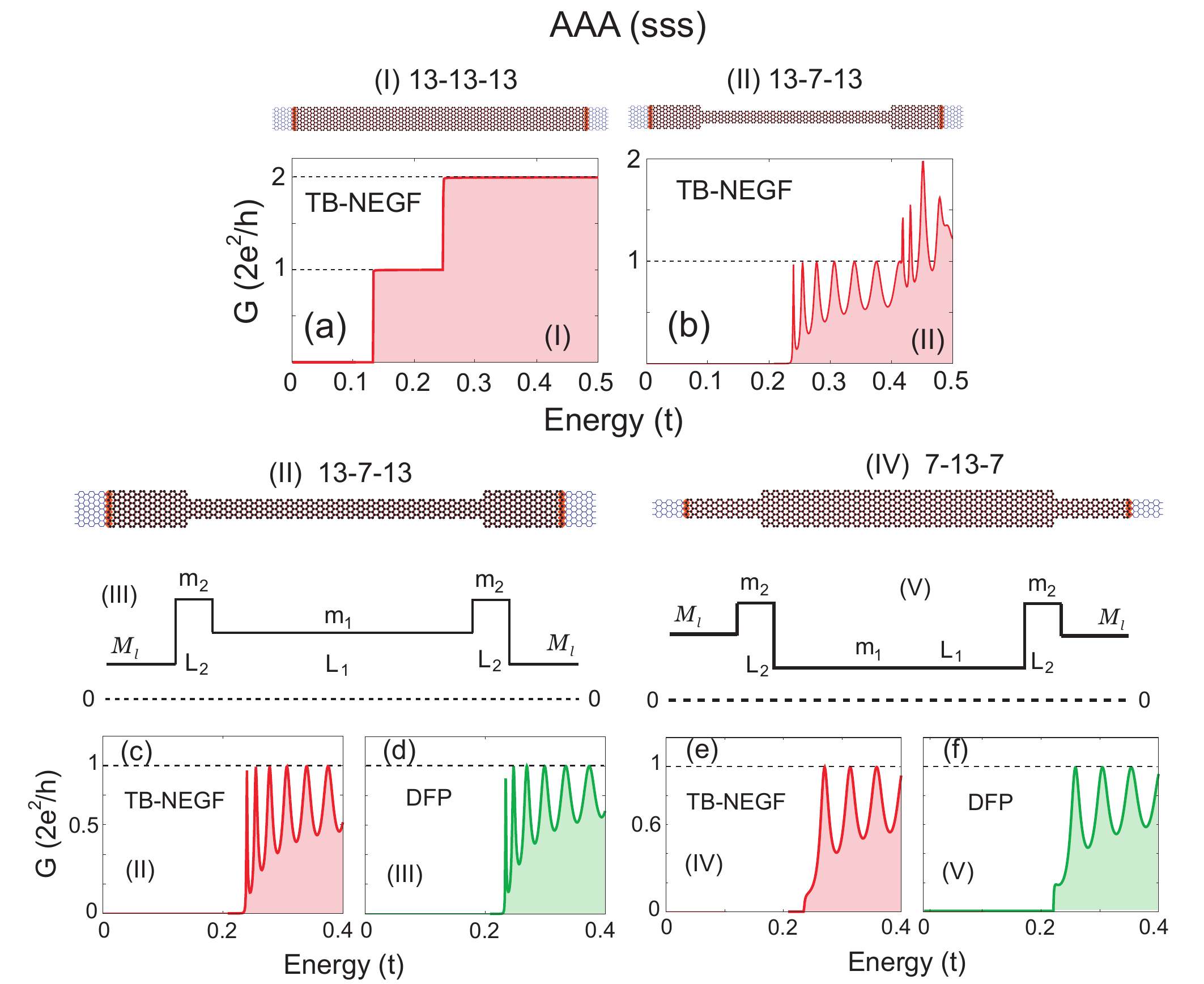}
\caption{{\bf Conductance quantization steps (a) for a uniform semiconducting armchair nanoribbon (I)
contrasted to Fabry-P\'{e}rot oscillations (b-f) of two 3-segment armchair GNRs [(II) and (IV)] with both a
semiconducting central constriction and semiconducting leads (13-7-13 and 7-13-7)}. 
For the systems shown here all the segments have armchair edge termination (hence, AAA), and all
have width corresponding to semiconducting GNRs [hence (sss)]. (III, V) Schematics of the mass barriers 
used in the DFP modeling, with the dashed line denoting the zero mass. The physics underlying such a junction 
is that of a massive relativistic Dirac fermion impinging upon the junction and performing multiple 
reflections (above $m_1 v_F^2$) within a particle box defined by the double-mass barrier. 
(c,e) TB-NEGF conductance as a function of the Fermi energy of the {\it massive\/} Dirac electrons in the leads.
(d) DFP conductance reproducing [in the energy range of the $1G_0$ step, see (b)] the TB-NEGF result in (c).
The mass-barrier parameters used in the DFP reproduction were $L_1=55 a_0$,  $m_1 v_F^2= 0.22 t$, $L_2=1 a_0$,
$m_2 v_F^2=0.5 t$. The mass of the electrons in the leads was ${\cal M}_l v_F^2=0.166 t$.
(f) DFP conductance reproducing the TB-NEGF result in (e).
The parameters used in the DFP reproduction were $L_1=53.6 a_0$,  $m_1 v_F^2= 0.166 t$, $L_2=1 a_0$,
$m_2 v_F^2=0.51 t$. The mass of the electrons in the leads was ${\cal M}_l v_F^2=0.22 t$.
$a_0=0.246$ nm is the graphene lattice constant; $t=2.7$ eV is the hopping parameter.}
\label{fig1}
\end{figure*}
~~~~~\\
~~~~~\\
~~~~~\\
~~~~~\\
~~~~~\\

\section{Results}

\subsection{Segmented Armchair GNRs: All-semiconducting}

Our results for a 3-segment (${\cal N}_1^W-{\cal N}_2^W-{\cal N}_1^W$, where ${\cal N}_1^W$ is the lead width 
and ${\cal N}_2^W$ is the width of the central segment) 
{\it all-semiconducting\/} aGNR, AAA (sss), are portrayed in Fig.\ \ref{fig1} [see schematic 
lattice diagrams in Figs.\ \ref{fig1}(I) and \ref{fig1}(II)].
A uniform semiconducting armchair GNR [see Fig.\ \ref{fig1}(I)] exhibits ballistic quantized-conductance steps 
[see Fig.\ \ref{fig1}(a)]. In contrast, conductance quantization is absent for a nonuniform 3-segment ($13-7-13$)
aGNR; see Figs.\ \ref{fig1}(b) $-$ \ref{fig1}(d). Here oscillations appear instead of quantized steps.
The first oscillation appears at an energy $\sim 0.22 t$ [Fig.\ \ref{fig1}(b)], which reflects the intrinsic
gap $\Delta/2$ of the semiconducting central segment belonging to the class II of aGNRs, specified 
\cite{fuji96,waka10,yann14} by a width ${\cal N}^W=3l+1$, $l=1,2,3,\ldots$. 
We recall that as a function of their width, ${\cal N}^W$, the armchair graphene nanoribbons fall into three 
classes: (I) ${\cal N}^W = 3l$ (semiconducting, $\Delta > 0$), (II) ${\cal N}^W = 3l+1$ (semiconducting, 
$\Delta > 0$), and (III) ${\cal N}^W = 3l+2$ (metallic, $\Delta = 0$), $l=1,2,3,\ldots$.

That the leads are semiconducting does not have any major effect.
This is due to the fact that ${\cal N}^W_2 < {\cal N}^W_1$, and as a result the energy gap $m_1 v_F^2$ of the 
central segment is larger than the energy gap ${\cal M}_l v_F^2$ of the semiconducting leads [see schematic
in Fig.\ \ref{fig1}(III)]. In the opposite case  (central segment wider than the leads), the energy gap 
of the semiconducting leads would have determined the onset of the conductance oscillations. 

The armchair GNR case with interchanged widths (i.e., $7-13-7$ instead of $13-7-13$) is portrayed in Figs.\ 
\ref{fig1}(e) $-$ \ref{fig1}(f). In this case the energy gap of the semiconducting leads (being the largest) 
determines the onset of the conductance oscillations. It is a testimonial of the consistency of our DFP method
that it can reproduce [see Fig.\ \ref{fig1}(d) and Fig.\ \ref{fig1}(f)] both the $13-7-13$ and 
$7-13-7$ TB-NEGF conductances; this is achieved with very similar sets of parameters taking into consideration
the central-segment-leads interchange. We note that the larger spacing between peaks (and also the smaller number
of peaks) in the $7-13-7$ case is due to the smaller mass of the central segment ($0.166t$ instead of $0.22t$). 

Further insight can be gained by an analysis of the discrete energies associated with the humps 
of the conductance oscillations in Fig.\ \ref{fig1}(c) and the resonant spikes in Fig.\ 
\ref{fig1}(e). Indeed a simplified approximation for the electron confinement in the continuum model consists 
in considering the graphene electrons as being trapped within a 1D infinite-mass square well (IMSW) of 
length $L_1$ (the mass terms are infinite outside the interval $L_1$ and the coupling to the leads 
vanishes). The discrete spectrum of the electrons in this case is given \cite{fiol96} by
\begin{equation} 
E_n=\sqrt{\hbar ^2 v_F^2 k_n^2+{\cal M}^2 v_F^4},
\label{erel} 
\end{equation}
where the wave numbers $k_n$ are solutions of the 
transcendental equation
\begin{equation}
\tan(k_n L_1) = -\hbar k_n/({\cal M} v_F).
\label{eimsw}
\end{equation}  
When ${\cal M}=0$ [massless Dirac-Weyl (DW) electrons], one 
finds for the spectrum of the IMSW model: 
\begin{equation}
E_n=(n+1/2) \pi \hbar v_F/L_1,
\label{imsw0}
\end{equation} 
with $n=0,1,2,\ldots$. 
\begin{equation} 
E_n-E_{n-1}=2E_0, \;\;\; n=1,2,3,\ldots, 
\label{edis}
\end{equation}
which is twice the energy 
\begin{equation}
E_0=\pi \hbar v_F/(2L_1)
\label{eoff} 
\end{equation}
of the lowest state.

As is well known, a constant energy separation of the intensity peaks, inversely proportional to the length
of the resonating cavity [here $L_1$, see Eqs. (\ref{edis}) and (\ref{eoff}) above] is the hallmark of the 
optical Fabry-P\'{e}rot, reflecting the linear energy dispersion of the photon in optics \cite{lipsbook} or 
a massless DW electron in graphene structures.\cite{lian01} Most revealing is the energy offset away form zero
of the first conductance peak, which equals exactly one-half of the constant energy separation between the peaks. 
In one-dimension, this is the hallmark of a massless fermion subject to an infinite-mass-barrier confinement 
\cite{fiol96}. Naturally, in the case of a semiconducting segment (see below), this equidistant 
behavior and $1/2$-offset of the conductance peaks do not apply; this case is accounted for by the 
Dirac-Fabry-P\'{e}rot model presented in Methods, and it is more general than the optical 
Fabry-P\'{e}rot theory associated with a photonic cavity. \cite{lipsbook} 

In the nonrelativistic limit, i.e., when $\hbar k_n \ll {\cal M} v_F$, one gets 
\begin{equation}
\tan(k_n L_1) \sim 0, 
\end{equation}
which yields the well known relations $k_n L_1 \sim n \pi$ and
\begin{equation} 
E_n \sim {\cal M}v_F^2 + n^2 \hbar^2 \pi^2/(2 {\cal M} L_1^2).
\label{ennr}
\end{equation}
For a massive relativistic electron, as is the case with the semiconducting aGNRs in this paper, one has to
numerically solve Eq.\ (\ref{eimsw}) and then substitute the corresponding value of $k_n$ in Einstein's energy 
relation given by Eq.\ (\ref{erel}). 

From an inspection of Fig.\ \ref{fig1}, one can conclude 
that the physics associated with the all-semiconducting AAA junction is that of multiple reflections of a 
{\it massive\/} relativistic Dirac fermion bouncing back and forth from the edges of a ``quantum box'' created by 
a double-mass barrier [see the schematic of the double-mass barrier in Figs.\ \ref{fig1}(III) and \ref{fig1}(V)]. 
In particular, to a good approximation the energies of the conductance oscillation peaks are given by the IMSW Eq.\
(\ref{eimsw}) with ${\cal M}v_F^2=m_1 v_F^2=0.22 t$ ($13-7-13$) or ${\cal M}v_F^2=m_1 v_F^2=0.166 t$
($7-13-7$). In this respect, the separation energy between successive peaks in Figs.\ \ref{fig1}(b), \ref{fig1}(c),
\ref{fig1}(e), and \ref{fig1}(f) is not a constant, unlike the case of an all-metallic junction \cite{lian01}
(or a photonic cavity \cite{lipsbook}). 

The conductance patterns in Figs.\ \ref{fig1}(d) and \ref{fig1}(f) correspond to the category {\it FP-B\/}
(see the introductory section). These patterns cannot be accounted for by the optical Fabry-P\'{e}rot theory, 
but they are well reproduced by the generalized Dirac-Fabry-P\'{e}rot model introduced by us in the Methods
section.

\newpage

{\centering\includegraphics[width=16cm]{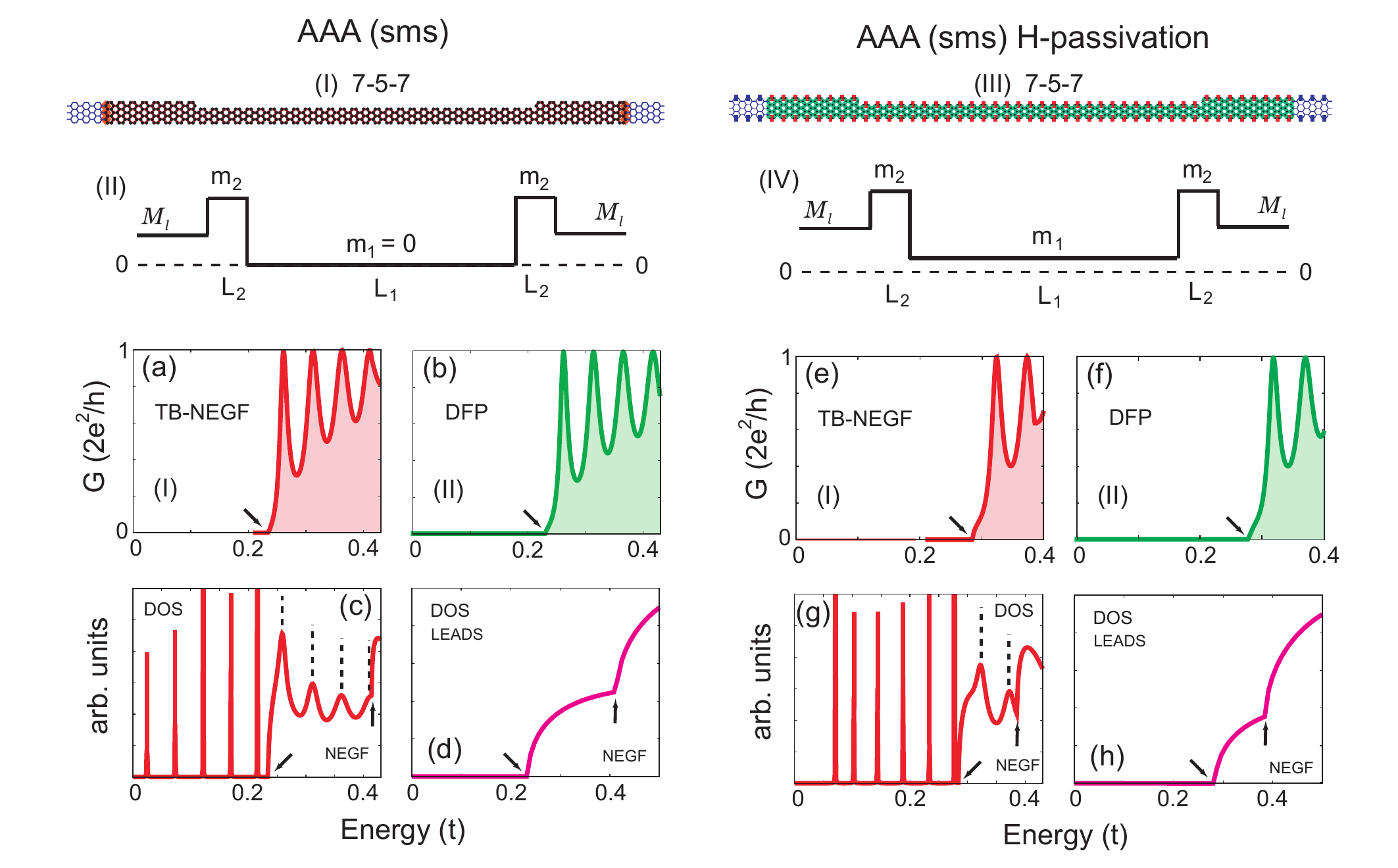}}
\newpage
\begin{figure*}
\caption{{\bf (Left) Conductance for a 3-segment nanoribbon, all segments having armchair edge termination 
(i.e. AAA), with a {\it metallic\/} (${\cal N}^W_2=5$) 
central constriction and semiconducting leads (${\cal N}^W_1=7$), hence the designation (sss)};
see schematic lattice diagram in (I). (II) Schematics of the mass barriers used in the DFP modeling. 
(a) TB-NEGF conductance as a function of the Fermi energy of the {\it massive\/} Dirac 
electrons in the leads. (b) DFP conductance reproducing the TB-NEGF result in (a). The mass-barrier parameters
used in the DFP reproduction were $L_1=60.4 a_0$,  $m_1 = 0$, $L_2=1 a_0$, $m_2 v_F^2=0.37 t$. The mass of the 
electrons in the leads was ${\cal M}_l v_F^2=0.23 t$. (c)-(d)  The total DOS of the junction and the density
of states in the isolated leads, respectively, according to the TB-NEGF calculations. The arrows indicate the
onset of the electronic bands in the leads. Note that the DOS in (c) reveal the existence of five sharp
electronic states below the onset (at $0.23 t \equiv {\cal M}_l v_F^2$) of the first band in the leads [see
(d)], which consequently do not generate any conductance resonances [see (a) and (b)]. Note further in (c) the 
equal energy spacing between the vertical lines [the five solid (red) and four dashed (black) ones] associated
with the resonances of a {\it massless\/} electron confined within the central metallic aGNR segment.\\
{\bf (Right) H-passivation effects in the conductance of a 3-segment armchair nanoribbon with a {\it metallic\/} 
(${\cal N}^W_2=5$) central constriction and semiconducting leads (${\cal N}^W_1=7$)}; see schematic 
lattice diagram in (III). Note that the nearest-neighbor C-C bonds at the armchair edges 
(thick red and blue lines) have hopping parameters $t^\prime=1.12 t$. (IV) Schematics of the
position-dependent mass field used in the DFP modeling. 
(e) TB-NEGF conductance as a function of the Fermi energy. (f) DFP conductance reproducing the 
TB-NEGF result in (e). The mass parameters used in the DFP reproduction were $L_1=59.5 a_0$,  $m_1 v_F^2= 0.05 t$, 
$L_2=1.5 a_0$, $m_2 v_F^2=0.30 t$. The mass of the electrons in the leads was ${\cal M}_l v_F^2=0.28t$. (g)-(h) 
The total DOS of the junction and the density of states in the isolated leads, respectively, according to the TB-NEGF
calculations. The arrows indicate the onset of the electronic bands in the leads; note the shifts from $0.23 t$ to
$0.28 t$ and from $0.42t$ to $0.38t$ for the onsets of the first and second bands, respectively, compared to the
case with $t^\prime=t$ in (d). Compared to left part of Fig.\ \ref{fig2}, the subtle modifications of mass 
parameters brought about by having $t^\prime =1.12 t$ result in having six sharp electronic states [see (g)] below 
the onset (at $0.28 t \equiv {\cal M}_l v_F^2$) of the first band in the leads [see (h)], which consequently do not 
generate any conductance resonances [see (e) and (f)]. In addition, within the energy range ($0$ to $0.4t$) shown in
(e) and (f), there are now only two conducting resonances, instead of three compared to (a) and (b). $a_0=0.246$ nm 
is the graphene lattice constant; $t=2.7$ eV is the graphene hopping parameter.}
\label{fig2}
\end{figure*}
\newpage

\subsection{Segmented Armchair GNRs: semiconducting-metallic-semiconducting.}
Our results for a 3-segment ($7-5-7$) {\it semiconducting-metallic-semiconducting\/} aGNR, AAA (sms), are
portrayed in Fig.\ \ref{fig2} [left, see schematic lattice diagram in Fig.\ \ref{fig2}(I)]. 
The first FP oscillation in the TB-NEGF conductance displayed in Fig.\ \ref{fig2}(a) appears at an energy 
$\sim 0.22 t$, which reflects the intrinsic gap $\Delta/2$ of the semiconducting leads (with ${\cal N}^W_1=7$). The
energy spacing between the peaks in Fig.\ \ref{fig2}(a) is constant in agreement with the metallic (massless DW
electrons) character of the central segment with ${\cal N}^W_2=5$. The TB-NEGF pattern in Fig.\ \ref{fig2}(a)
corresponds to the Fabry-P\'{e}rot category {\it FP-A2\/}. As seen from Fig.\ \ref{fig2}(b), our generalized 
Dirac-Fabry-P\'{e}rot theory is again capable of faithfully reproducing this behavior.

A deeper understanding of the AAA (sms) case can be gained via an inspection of the density of states (DOS) plotted 
in Fig.\ \ref{fig2}(c) for the total segmented aGNR (central segment plus leads) and in Fig.\ \ref{fig2}(d) for the  
the isolated leads. In Fig.\ \ref{fig2}(c), nine equidistant resonance lines are seen. Their energies are close to
those resulting from the IMSW Eq.\ (\ref{imsw0}) (with $L_1=60.4 a_0$, see the caption of Fig.\ \ref{fig2}) for a
massless DW electron. Out of these nine resonances, the first five do not conduct [compare Figs.\ \ref{fig2}(a) and
\ref{fig2}(c)] because their energies are lower than the minimum energy (i.e., $\Delta/2={\cal M}_l v_F^2 \sim
0.23 t$) of the incoming electrons in the leads [see the onset of the first band (marked by an arrow) in the
DOS curve displayed in Fig.\ \ref{fig2}(d)]. 

\subsection{Segmented Armchair GNRs: Effects of hydrogen passivation.}
As shown in Refs.\ \citenum{cohe06,wang07}, a detailed description of hygrogen passivation requires that the
hopping parameters $t^\prime$ for the nearest-neighbor C-C bonds at the armchair edges be given by $t^\prime=1.12t$.  
Taking this modification into account, our results for a 3-segment {\it semiconducting-metallic-semiconducting\/} 
aGNR are portrayed in Fig.\ \ref{fig2} [right, see schematic lattice diagram in Fig.\ \ref{fig2}(III)]; 
this lattice configuration is denoted as "AAA (sms) H-passivation." The first FP oscillation in the 
TB-NEGF conductance displayed in Fig.\ \ref{fig2}(e) appears at an energy $\sim 0.28 t$, which reflects the 
intrinsic gap $\Delta/2$ of the properly passivated semiconducting leads (with ${\cal N}^W_1=7$). The energy spacing 
between the peaks in Fig.\ \ref{fig2}(e) is slightly away from being constant in agreement with the small mass 
$m_1 v_F^2=0.05t$ acquired by the central segment with ${\cal N}^W_2=5$, due to taking $t^\prime=1.12t$. As seen from
Fig.\ \ref{fig2}(f), our generalized Dirac-Fabry-P\'{e}rot theory is again capable of faithfully reproducing
this behavior.

A deeper understanding of the AAA (sms)-H-passivation case can be gained via an inspection of the DOS plotted in 
Fig.\ \ref{fig2}(g) for the total segmented aGNR (central segment plus leads) and in Fig.\ \ref{fig2}(h) for the 
isolated leads. In Fig.\ \ref{fig2}(g), eight (almost, but not exactly, equidistant) resonance lines are
seen. Their energies are close to those resulting from the IMSW Eq.\ (\ref{eimsw}) 
(with $L_1=59.5 a_0$ and $m_1 v_F^2=0.05t$; see the caption of Fig.\ \ref{fig2}) for a Dirac electron with a small 
mass. Out of these eight resonances, the first six do not conduct [compare Figs.\ \ref{fig2}(e) and \ref{fig2}(g)] 
because their energies are lower than the minimum energy (i.e., $\Delta/2={\cal M}_l v_F^2 \sim 0.28 t$) of the 
incoming electrons in the leads [see the onset of the first band (marked by an arrow) in the DOS curve displayed in 
Fig.\ \ref{fig2}(h)]. 
From the above we conclude that hydrogen passivation of the aGNR resulted in a small shift of the location of
the states, and opening of a small gap for the central metallic narrower (with a width of $N_2^W=5$) segment, but did
not modify the conductance record in any qualitative way. Moreover, the passivation effect can be faithfully
captured by the Dirac FP model by a small readjustment of the model parameters. 

\begin{figure}
\centering\includegraphics[width=11.5cm]{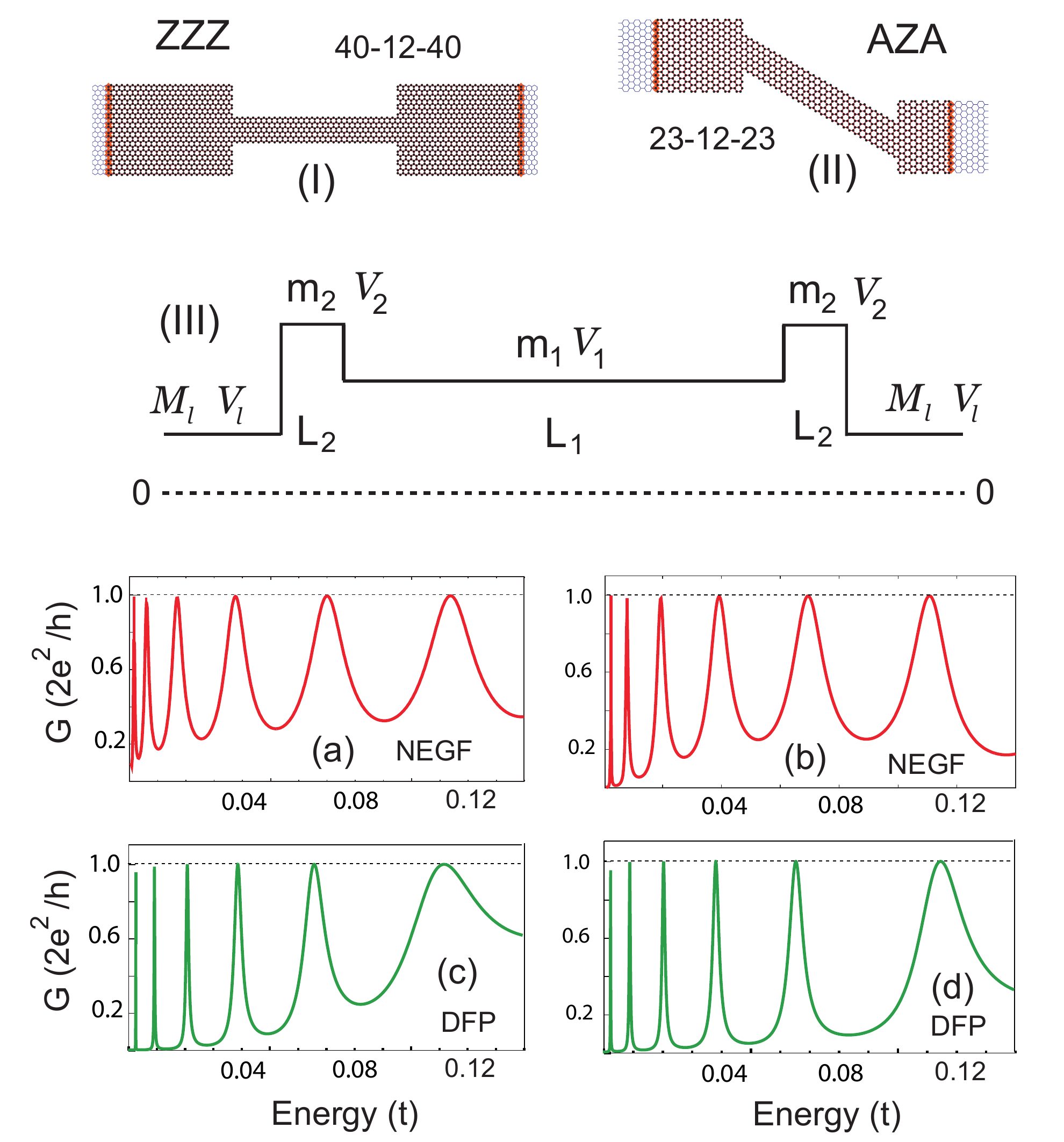}
\caption{{\bf Conductance for ZZZ (all-zigzag edge termination, left column) and AZA (armchair-zigzag-armchair
edge termination, right column) segmented nanoribbon junctions.} See corresponding lattice diagrams in (I) and (II). 
The 3-segment GNRs are denoted as ${\cal N}^W_1 - {\cal N}^W_2 - {\cal N}^W_1$, with ${\cal N}^W_i$ ($i=1,2$) 
being the number of carbon atoms specifying the width of the ribbon segments. The armchair leads in 
the AZA junction are metallic (${\cal N}^W_1=23$, class III aGNR).  
(a)-(b) TB-NEGF conductance for the ZZZ and AZA junction, respectively. 
(c) DFP conductance reproducing the TB-NEGF result in (a) for the ZZZ junction.
(d) DFP conductance reproducing the TB-NEGF result in (b) for the AZA junction.  
In spite of the different edge morphology, the Fabry-P\'{e}rot patterns in (a) and (b) are very similar.
The central zigzag segment controls the Fabry-P\'{e}rot patterns. According to the continuum DFP analysis,
the physics underlying such patterns is that of a massive nonrelativistic Schr\"{o}dinger fermionic carrier 
performing multiple reflections within a cavity defined by a double-mass barrier [see diagram in (III)], but 
with the additional feature that $V_1=-m_1 v_F^2$ and $V_l=-{\cal M}_l v_F^2$ are also considered for segments
or leads with zigzag edge terminations (see text for details). The mass and $V_i$ parameters used in the DFP 
calculations were $L_1=30 a_0$,  $m_1 v_F^2=2.23 t - c E t$, with $c=7.3$, $V_1=-m_1 v_F^2$, $L_2=1.1 a_0$, 
$m_2 v_F^2=0.38 t$, $V_2=-m_2 v_F^2/3$, ${\cal M}_l v_F^2= 2.30 t$, $V_l=-{\cal M}_l v_F^2$ in (c) and 
$L_1=29.1 a_0$,  $m_1 v_F^2=2.65 t - c E t$, with $c=8.4$, $V_1=-m_1 v_F^2$, $L_2=1.0 a_0$, $m_2 v_F^2=0.30 t$, 
$V_2=-m_2 v_F^2$, ${\cal M}_l=0$, $V_l=0$ in (d). ${\cal M}_l$ and $V_l$ denote parameters of the leads.
$E$ is the energy in units of $t$. $a_0=0.246$ nm is the graphene lattice constant; $t=2.7$ eV is the hopping 
parameter.}
\label{fig3}
\end{figure}

\subsection{All-zigzag segmented GNRs.}
It is interesting to investigate the sensitivity of the interference features on the edge morphology.
We show in this section that the relativistic transport treatment applied to segmented armchaie GNRs does not 
maintain for the case of a nanoribbon segment with zigzag edge terminations. In fact zigzag GNR (zGNR) segments
exhibit properties akin to the well-known transport in usual semiconductors, i.e., their 
excitations are governed by the nonrelativistic Schr\"{o}dinger equation.

Before discussing segmented GNRs with zigzag edge terminations, we remark that such GNRs with uniform width
exhibit stepwise quantization of the conductance, similar to the case of a uniform 
armchair-edge-terminated GNR [see Fig.\ \ref{fig1}(a)].

In Fig.\ \ref{fig3}(a), we display the conductance in a three-segment junction [see lattice schematic in Fig.\ 
\ref{fig3}(I)] when all three segments have zigzag edge terminations (denoted as ZZZ), but the central one is
narrower than the lead segments. The main finding is that the central segment behaves again as a resonant
cavity that yields an oscillatory conductance pattern where the peak spacings are unequal [Fig.\ \ref{fig3}(a)]. 
This feature, which deviates from the optical Fabry-P\'{e}rot behavior, appeared also in the DFP patterns for
a three-segment armchair junction whose central segment was semiconducting, albeit with a different dependence
on $L$ [see Figs.\ \ref{fig1} and Eq.\ (\ref{eimsw})]. Moreover, from a set of
systematic calculations (not shown) employing different lengths and widths, we found that the energy of
the resonant levels in zGNR segments varies on the average as $\sim (n/L)^2$, where the integer $n$ counts the 
resonances and $L$ indicates the length of the central segment. However, a determining difference with
the armchair GNR case in Fig.\ \ref{fig1} is the vanishing of the valence-to-conductance gap
in the zigzag case of Fig.\ \ref{fig3}(a). It is well known that the above features 
are associated with resonant transport of electronic excitations that obey the nonrelativistic second-order 
Schr\"{o}dinger equation.

Naturally, one could formulate a continuum transport theory based on transfer matrices (see Methods) 
that use the 1D Schr\"{o}dinger equation instead of the generalized Dirac Eq.\ (\ref{direq}).  
Such a Schr\"{o}dinger-equation continuum approach, however, is unable to describe mixed armchair-zigzag
interfaces (see below), where the electron transits between two extreme regimes, i.e., an 
ultrarelativistic (i.e., including the limit of vanishing carrier mass) Dirac regime (armchair segment) and a 
nonrelativistic Schr\"{o}dinger regime (zigzag segment). We have thus been led to adopt the same Dirac-type 
transfer-matrix approach as with the armchair GNRs, but with nonvanishing potentials $V=\mp {\cal M} v_F^2$. 
This amounts to shifts (in opposite senses) of the energy scales for particle and hole excitations, respectively,
and it yields the desired vanishing value for the valence-to-conduction gap of zigzag GNRs.

The calculated DFP conductance that reproduces well the TB-NEGF result for the ZZZ junction [Fig.\
\ref{fig3}(a)] is displayed in Fig.\ \ref{fig3}(c); the parameters used in the DFP calculation 
are given in the caption of Fig.\ \ref{fig3}. We note that the carrier mass ($m_1$) in the central zigzag segment 
exhibits an energy dependence. This is similar to a well known effect (due to nonparabolicity in the $E-k$ 
dispersion) in the transport theory of usual semiconductors \cite{melc93}. We further note that the average 
mass associated with a zigzag segment is an order of magnitude larger than that found for semiconducting 
armchair segments of similar width (see caption in Fig.\ \ref{fig1}), and this yields energy 
levels $\sim (n/L)^2$ close to the nonrelativistic limit [see Eq.\ (\ref{ennr})]. We note that the FP pattern 
of the ZZZ junction belongs to the category {\it FP-C\/}. 

\subsection{Mixed armchair-zigzag-armchair segmented GNRs.}
Fig.\ \ref{fig3} (right column) presents an example of a mixed armchair-zigzag-armchair (AZA) junction, where 
the central segment has again zigzag edge terminations [see lattice schematic in Fig.\ \ref{fig3}(II)]. The 
corresponding TB-NEGF conductance is displayed in Fig.\ \ref{fig3}(b). In spite of the different morphology 
of the edges between the leads (armchair) and the central segment (zigzag), the conductance profile of
the AZA junction [Fig.\ \ref{fig3}(b)] is very similar to that of the ZZZ junction [Fig.\ \ref{fig3}(a)].
This means that the characteristics of the transport are determined mainly by the central segment, with the
left and right leads, whether zigzag or armchair, acting as reservoirs supplying the impinging electrons. 

The DFP result reproducing the TB-NEGF conductance is displayed in Fig.\ \ref{fig3}(d), and the parameters
used are given in the caption. We stress that the mixed AZA junction represents a rather unusual physical
regime, where an ultrarelativistic Dirac-Weyl massless charge carrier (due to the metallic armchair GNRs 
in the leads) transits to a nonrelativistic massive Schr\"{o}dinger electron in the central segment. We note that 
the FP pattern of the AZA junction belongs to the {\it FP-C\/} caregory.

\begin{figure}
\centering\includegraphics[width=13.5cm]{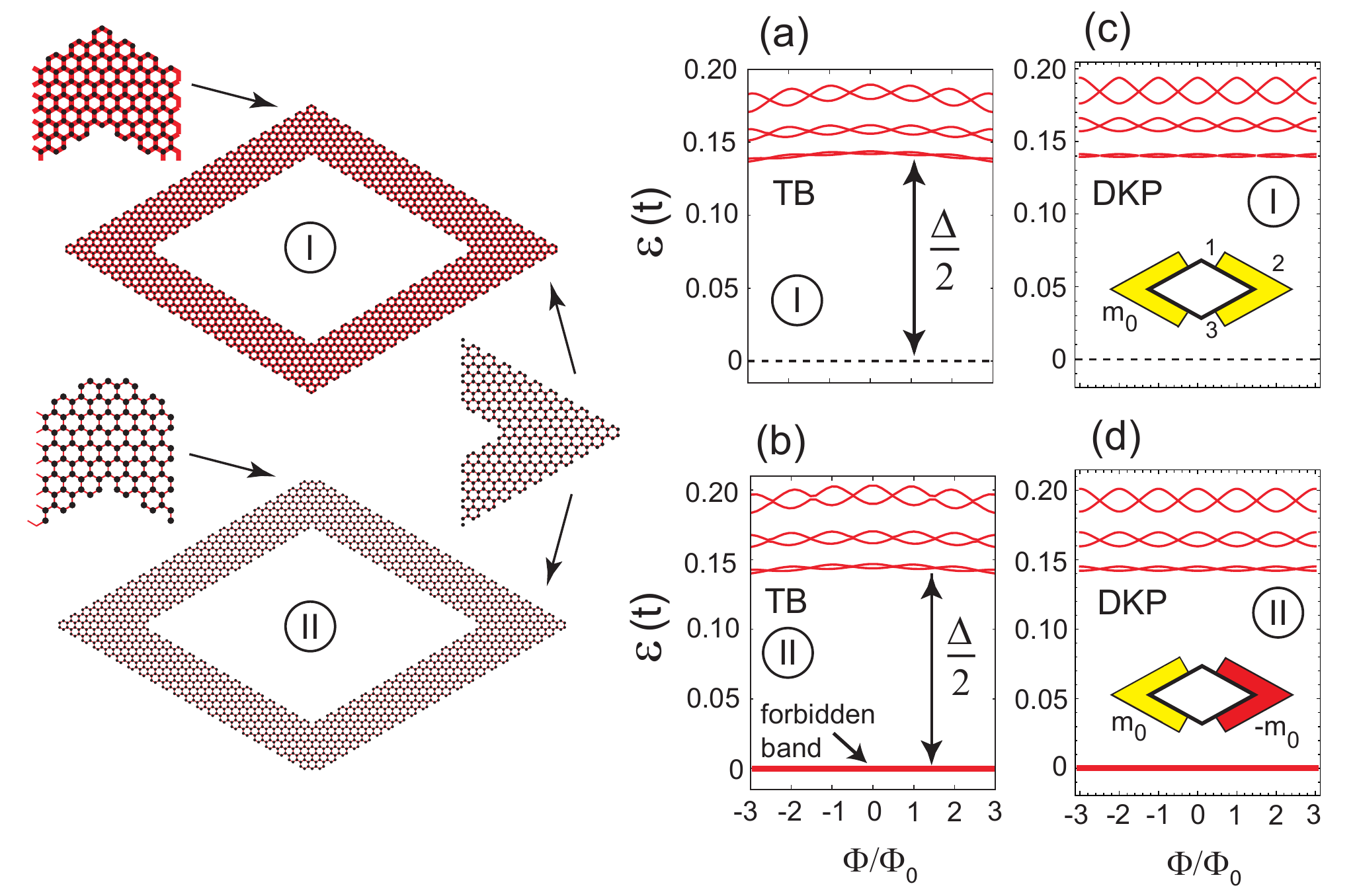}
\caption{
{\bf Aharonov-Bohm spectra for rhombic armchair graphene rings;  the two rings that we consider (I and II on
the left) show different atomic arrangement at the top and bottom corners.}
(a) Tight-binding spectrum for a nanoring with type-I corners and width ${\cal N}^W=12$. (b) TB spectrum for a 
nanoring with type-II corners and the same width ${\cal N}^W=12$. These armchair graphene rings are
semiconducting (type-I) and metallic (type-II).  
The three (four)  lowest-in-energy two-membered bands are shown. The hole states (with 
$\varepsilon < 0$, not shown) are symmetric to the particle states (with $\varepsilon > 0$). 
(c,d) DKP spectra reproducing the TB ones in (a) and (b), respectively. Insets in (c) and (d): schematics of 
the Higgs fields (position-dependent mass) $\phi(x)$ employed in the DKP modeling. $\phi(x)$ is 
approximated by steplike functions $m_i^{(n)}$; $i$ counts the three regions of each half of the rhombus
($L_1^{(n)}=L_3^{(n)}=a$ and $L_2^{(n)}=b$), and $n$ ($n=1,2$). The non-zero (constant) variable-mass values of
$\phi(x)$ are indicated by yellow (red) color when positive (negative). 
The parameters used in the DKP modeling are: (c) $a=1.3 a_0$, $b=66 a_0$, $m_1^{(n)}=m_3^{(n)}=0$, 
$m_2^{(n)}=m_0=0.15 t/v_F^2$ [see corresponding schematic inset in (c)] 
and (d)  $a=6.5 a_0$, $b=55 a_0$, $m_1^{(n)}=m_3^{(n)}=0$, $m_2^{(n)}=(-1)^n m_0$ with 
$m_0=0.15 t/v_F^2$ [see schematic inset in (d)]. 
Note the two-membered braided bands and the ``forbidden'' band [within the gap, in (b) and (d)]. The twofold 
forbidden band with $\epsilon \sim 0$ appears as a straight line due to the very small amplitude oscillations 
of its two members. $a_0=0.246$ nm is the graphene lattice constant and $t=2.7$ eV is the hopping parameter.
The edge terminations of both the inside and outside sides of the ring are armchair.}
\label{fig44}
\end{figure}

\subsection{Aharonov-Bohm spectra of rhombic graphene rings.}

The energy of a particle (with onedimensional momentum $p_x$) is given by the Einstein relativistic relation 
$E=\sqrt{ (p_x v_F)^2+({\cal M} v_F^2)^2 }$, where ${\cal M}$ is the rest mass. As aforementioned, in armchair 
graphene ribbons, the mass parameter is related to the particle-hole energy gap, $\Delta$, as 
${\cal M}=\Delta/(2 v_F^2)$. In relativistic quantum field theory, the mass of 
elementary particles is imparted through interaction with a scalar field known as the Higgs field.
Accordingly, the mass ${\cal M}$ is replaced by a position-dependent Higgs field 
$\phi(x) \equiv m(x)$, to which the relativistic fermionic field $\Psi(x)$ couples through the Yukawa 
Lagrangian \cite{mapa90_1,mapa90_2,roma13} ${\cal L}_Y = -\phi \Psi^\dagger \beta \Psi$ ($\beta$ being a Pauli 
matrix). In the elementary-particles Standard Model, \cite{griffbook} such coupling 
is responsible for the masses of quarks and leptons. For $\phi(x) \equiv \phi_0$ (constant)
${\cal M} v_F^2=\phi_0$, and the massive fermion Dirac theory is recovered.

We exploit the generalized Dirac physics governed by a total Lagrangian density ${\cal L}=
{\cal L}_f + {\cal L}_\phi$, where the fermionic part is given by 
\begin{equation}
{\cal L}_f = - i \hbar \Psi^\dagger \frac{\partial}{\partial t} \Psi
- i \hbar v_F \Psi^\dagger \alpha \frac{\partial}{\partial x} \Psi +{\cal L}_Y ,
\label{lagrf}
\end{equation}
and the scalar-field part has the form
\begin{equation}
{\cal L}_\phi =  - \frac{1}{2} (\frac{\partial \phi}{\partial x})^2 - 
\frac{\xi}{4} (\phi^2 - \phi_0^2)^2,
\label{lagrphi}
\end{equation}
with the potential $V(\phi)$ (second term) assumed to have a double-well $\phi^4$ form;
$\xi$ and $\phi_0$ are constants.

Henceforth, the Dirac equation (see Methods) is generalized as
\begin{equation}
E \Psi + i \hbar v_F \alpha \frac{\partial \Psi}{\partial x} - \beta \phi(x) \Psi=0.
\label{direq1}
\end{equation} 
In one dimension, the fermion field is a two-component spinor $\Psi = (\psi_u, \psi_l)^T$;
$u$ and $l$ stand, respectively, for the upper and lower component and
$\alpha$ and $\beta$ can be any two of the three Pauli matrices. 

\begin{figure}[t]
\centering\includegraphics[width=12.5cm]{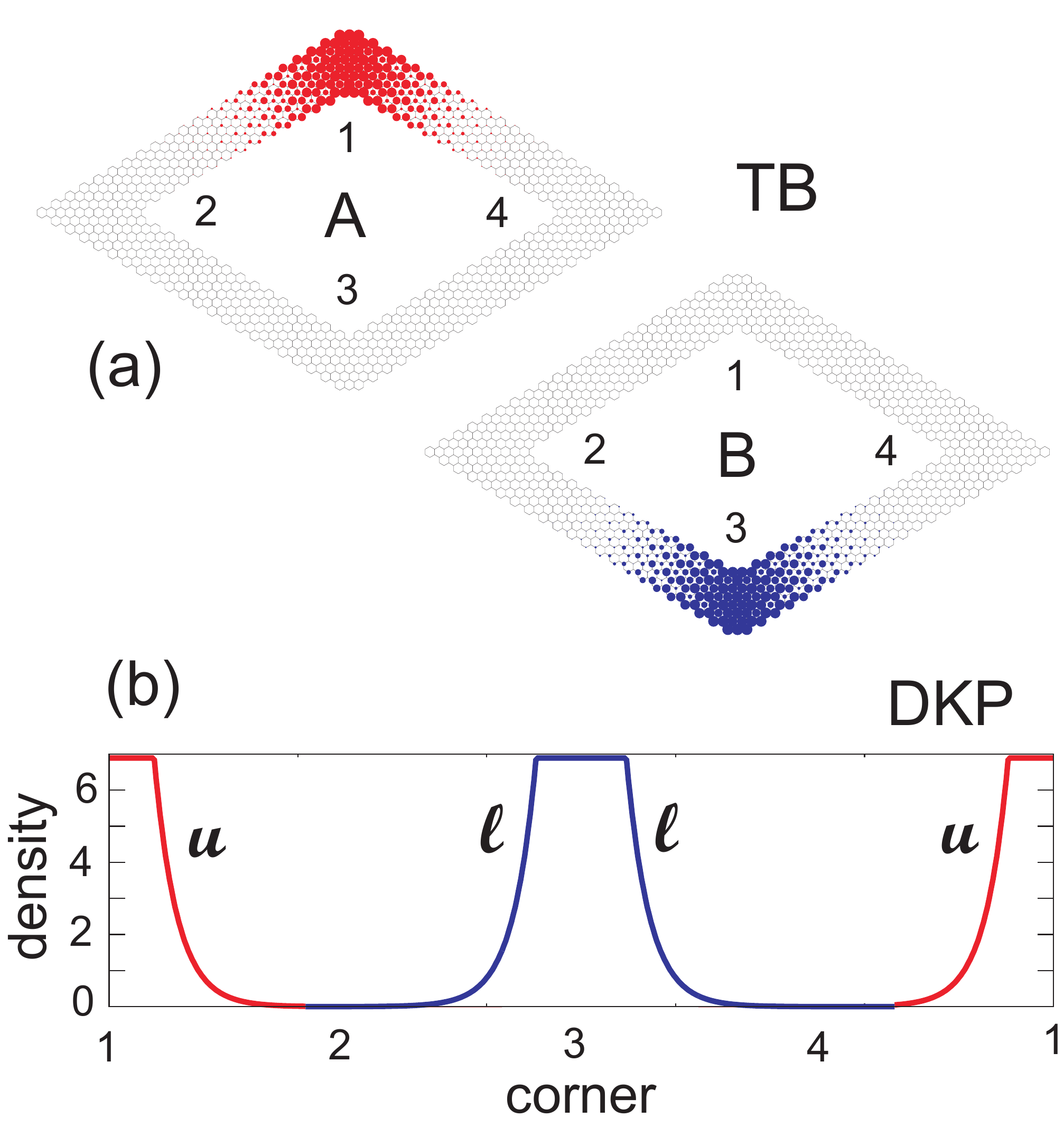}
\caption{
{\bf Wave functions for an excitation belonging to the ``forbidden'' solitonic  band.} 
(a) $A$-sublattice (red) and $B$-sublattice (blue) components of the TB state with energy
$\varepsilon=0.280 \times 10^{-4}t$ at $\Phi=0$, belonging to the forbidden solitonic band
of the type-II nanoring with ${\cal N}^W=12$ [see Fig.\ \ref{fig44}(b)]. 
(b) Upper (red) and lower (blue) spinor components for the corresponding state  
(forbidden band) according to the DKP spectrum [see Fig.\ \ref{fig44}(d)], reproducing 
the TB behavior of the type-II nanoring with ${\cal N}^W=12$ ($m_0=0.15 t/v_F^2$).
The TB and DKP wave functions for all states of the solitonic band are similar to those
displayed here. The wave functions here represent a pair of solitons. For contrast, see 
Fig.\ 10 in Ref.\ \citenum{roma13} which portrays schematically the spinor $\Psi_S$ for a 
{\it single\/} fermionic soliton attached to a Higgs field with a smooth kink-soliton 
analytic shape $\phi_k(x)= \phi_0 \tanh \left( \sqrt{\xi/2} \phi_0 x \right)$. 
$\phi_k(x)$ is a solution \cite{jack81,roma13} of the Lagrangian in Eq.\ (\ref{lagrphi}). 
DKP densities in units of $10^{-3}/a_0$.
}   
\label{fig55}
\end{figure}

A graphene polygonal ring can be viewed as made of connected graphene-nanoribbon fragments (here we consider
aGNRs). The excitations of an infinite aGNR are described by the 1D massive Dirac 
equation, see Eq.\ (\ref{direq1}) with $\alpha=\sigma_2$, $\beta=\sigma_1$, and 
$\phi(x) \equiv \phi_0 = \Delta/2 \equiv |t_1-t_2|$. The two (in general) unequal hopping parameters 
$t_1$ and $t_2$ are associated with an effective 1D tight-binding problem) 
and are given \cite{zhen07} by 
$t_1=-2 t \cos[p \pi/({\cal N}^W+1)]$, $p=1,2,\ldots,{\cal N}^W$ and $t_2=-t$; ${\cal N}^W$ is 
the number of carbon atoms specifying the width of the nanoribbon and $t=2.7$ eV is the hopping 
parameter for 2D graphene. The effective \cite{zhen07} TB Hamiltonian of an aGNR has a 
form similar to that used in {\it trans}-polyacetylene (a single chain of carbon atoms). In 
{\it trans}-polyacetylene, the inequality of $t_1$ and $t_2$ (referred to as dimerization) is a 
consequence of the aforementioned Peierls distortion induced by the electron-phonon coupling. For
an armchair graphene ring, this inequality is a topological effect associated with 
the geometry of the edge and the width of the ribbon. We recall that as a function of their width, 
${\cal N}^W$, the armchair graphene nanoribbons fall into three classes: (I) ${\cal N}^W=3l$ 
(semiconducting, $\Delta>0$), (II) ${\cal N}^W=3l+1$ (semiconducting, $\Delta>0$), and (III) 
${\cal N}^W=3l+2$ (metallic $\Delta=0$), $l=1,2,3,\ldots$.  

We adapt the ``crystal'' approach \cite{imry83} to the Aharonov-Bohm (AB) effect, and introduce a virtual
Dirac-Kronig-Penney \cite{mcke87} (DKP) relativistic superlattice (see Methods). Charged 
fermions in a perpendicular magnetic field circulating around the ring behave like electrons in a 
spatially periodic structure (period ${\cal D}$) with the magnetic flux $\Phi/\Phi_0$ ($\Phi_0=hc/e$) 
playing the role of the Bloch wave vector $k$, i.e., $2 \pi \Phi/\Phi_0=k {\cal D}$ [see the cosine 
term in Eq.\ (\ref{disrel})]. 

Naturally, nanorings with arms made of nanoribbon segments belonging to the semiconducting classes 
may be expected to exhibit a particle-hole gap (particle-antiparticle gap in RQF theory). 
Indeed this is found for a rhombic armchair graphene ring (AGR) [see gap $\Delta$ in Fig.\ \ref{fig44}(a)]
with a width of ${\cal N}^W=12$ carbon atoms having type-I corners. Suprisingly, a rhombic armchair graphene 
nanoring of the same width ${\cal N}^W=12$, but having corners of type-II, demonstrates a different behavior, 
showing a ``forbidden'' band (with $\epsilon \sim 0$) in the middle of the gap region [see Fig.\ \ref{fig44}(b)]. 

This behavior of rhombic armchair graphene rings with type-II corners can be explained through analogies with 
RQF theoretical models, describing single zero-energy fermionic solitons with fractional charge 
\cite{jare76,jack81} or their modifications when forming soliton/anti-soliton systems. \cite{jakl83,jack81} 
(A solution of the equation of motion corresponding to Eq.\ (\ref{lagrphi}), is a $Z_2$
kink soliton, $\phi_k(x)$. The solution of Eq.\ (\ref{direq1}) with $\phi=\phi_k(x)$ is the 
fermionic soliton.) We model the rhombic ring with the use of a continuous 1D Kronig-Penney 
\cite{mcke87} model (see Methods) based on the generalized Dirac equation 
(\ref{direq1}), allowing variation of the scalar 
field $\phi(x)$ along the ring's arms. We find that the DKP model reproduces [see Fig.\ 
\ref{fig44}(d)] the spectrum of the type-II rhombic ring (including the forbidden band) 
when considering alternating masses $\pm m_0$ associated with each half of the ring 
[see inset in Fig.\ \ref{fig44}(d)].

In analogy with the physics of {\it trans\/}-polyacetylene (see remarks in the introductory section), 
the positive and negative masses correspond 
to two degenerate domains associated with the two possible dimerization patterns \cite{jack81,heeg88} 
$\ldots -t_1-t_2-t_1-t_2- \ldots$ and $\ldots -t_2-t_1-t_2-t_1- \ldots$, which are possible in a 
single-atom chain. The transition zones between the two domains (here two of the four corners of the rhombic 
ring) are referred to as the domain walls.

For a single soliton, a (precise) zero-energy fermionic excitation emerges, 
localized at the domain wall. In the case of soliton-antisoliton pairs, paired energy levels with
small positive and negative values appear within the gap. The TB spectrum 
in Fig.\ \ref{fig44}(b) exhibits a forbidden band of two paired $+/-$ 
levels, a property fully reproduced by the DKP model that employs two alternating mass domains 
[Fig.\ \ref{fig44}(d). The twofold forbidden band with $\epsilon \sim 0$ appears as a straight line due to the 
very small amplitude oscillations of its two members. 
$a_0=0.246$ nm is the graphene lattice constant and $t=2.7$ eV is the hopping parameter.

The strong localization of a fraction of a fermion at the domain walls (two of the rhombus' corners), 
characteristic of fermionic solitons \cite{jack81} and of soliton/anti-soliton 
pairs, \cite{jakl83} is clearly seen in the TB density distributions (modulus of single-particle 
wave functions) displayed in Fig.\ \ref{fig55}(a). The TB $A$ ($B$) sublattice component of the tight-binding 
wave functions localizes at the odd numbered corners. These alternating localization patterns are 
faithfully reproduced [see Fig.\ \ref{fig55}(b)] by the upper, $\psi_u$, and lower, $\psi_l$, spinor 
components of the continuum DKP model. The soliton-antisoliton pair in Fig.\ \ref{fig55}(b) generates an $e/2$ 
charge fractionization at each of the odd-numbered corners, which is similar to the 
$e/2$ fractionization familiar from polyacetylene. 

The absence of a forbidden band (i.e., solitonic excitations within the gap) in the spectrum of the 
type-I rhombic nanorings [see Fig.\ \ref{fig44}(a)] indicates that the 
corners in this case do not induce an alternation between the two equivalent dimerized domains 
(represented by $\pm m_0$ in the DKP model). Here the corners do not act as topological domain walls. 
Nevertheless, direct correspondence between the TB and DKP spectra is 
achieved here too by using a variable Higgs field defined as $\phi(x)=m^{(n)}_i(x)$ with 
$m_1^{(n)}=m_3^{(n)}= 0$ and $m_2^{(n)}=m_0=0.15 t/v_F^2$ [see the schematic inset in Fig.\ \ref{fig44}(c) 
and the DKP spectrum plotted in the same figure]. 

\section{Discussion}

In this paper we focused on manifestations of relativistic and/or nonrelativistic quantum behavior explored
through theoretical considerations of transport in graphene nanostructures and spectral and topological
effects in graphene nanorings in the presence of magnetic fields. In particular, we investigated the emergence
of new behavior of electrons in atomically precise segmented graphene nanoribbons (GNRs) with different edge
terminations (armchair, zigzag and mixed ones), and in graphene rings. To these aims we have employed
tight-binding calculations of electronic states with and without applied magnetic fields, the non-equilibrium
Green's function transport theory, and a newly developed Dirac continuum model that absorbs the
valence-to-conductance energy gaps as position-dependent masses, including topological-in-origin mass-barriers
at the contacts between segments. 

The electronic conductance has been found to exhibit Fabry-P\'{e}rot oscillations, or resonant tunneling,
associated with partial confinement and formation of a quantum box (resonant cavity) in the junction. Along
with the familiar optical FP oscillations, exhibiting equal spacing between neighboring peaks, that we find
for massless electrons in GNRs with metallic armchair central segments, we find other FP categories that 
differ from the optical one. In particular, our calculations reveal: (a) A massive relativistic FP pattern
exhibiting a valence-to-conduction gap and unequal peak spacings. This pattern is associated with
semiconducting armchair nanoribbon central segments, irrespective of whether the armchair leads are matallic
or semiconducting. (b) A massive non-relativistic FP pattern with $1/L^2$ peak spacings, but with a vanishing
valence-to-conduction gap. This pattern is the one expected for the carriers in usual semiconductors described
by the (nonrelativistic) Schr\"{o}dinger equation, and it is associated with zigzag nanoribbon central 
segments, regardless of whether zigzag or metallic armchair leads are used. 

Perfect quantized-conductance flat steps were found only for uniform GNRs. In the absence of extraneous
factors, like disorder, in our theoretical model, the deviations from the perfect quantized-conductance steps
were unexpected. However, this aforementioned behavior obtained through TB-NEGF calculations is well accounted
for by a 1D contimuum fermionic Dirac-Fabry-P\'{e}rot interference theory (see Methods). This approach employs
an effective position-dependent mass term in the Dirac Hamiltonian to absorb the finite-width
(valence-to-conduction) gap in armchair nanoribbon segments, as well as the barriers at the interfaces between
nanoribbon segments forming a junction. For zigzag nanoribbon segments the mass term in the Dirac equation
reflects the nonrelativistic Schrodinger-type behavior of the excitations. The carrier mass in
zigzag-terminated GNR segments is much larger than the particle mass in semiconducting armchair-terminated GNR
segments. Furthermore in the zigzag GNR segments (which are always characterized by a vanishing
valence-to-conduction energy gap), the mass corresponds simply to the carrier mass. In the armchair GNR
segments, the carrier mass endows (in addition) the segment with a valence-to-conduction energy gap, according
to Einstein's relativistic energy relation [see Eq.\ (1)].

We concluded with a brief discussion of the physics of electrons in segmented polygonal rings, which may be
regarded as constructed by connecting GNR segments. Evaluation of the electronic states in a rhombic
graphene nanoring under the influence of an applied magnetic field in the Aharonov-Bohm regime, and their
analysis with the use of a relativistic quantum-field theoretical model, unveils development of a
topological-in-origin zero-energy soliton state and charge fractionization.

The above findings point to a most fundamental underlying physics, namely that the topology of disruptions of
the regular honeycomb lattice (e.g., variable width segments, corners, edges) generates a scalar-potential
field (position-dependent mass, identified \cite{roma13,yann14} also as a Higgs-type field), which when 
integrated into a generalized Dirac equation for the electrons provides a unifying framework for the analysis 
of transport processes through graphene segmented junctions and the nature of electronic states in  
graphene nanorings.

With growing activities and further improvements in the areas of bottom-up fabrication and manipulation
of atomically precise \cite{ruff10,fuhr10,huan12,derl13,nari14,hart14,sini14} 
graphene nanostructures and the anticipated measurement
of conductance through them, the above findings could serve as impetus and implements 
aiding the design and interpretation of future experiments.

~~~~\\
\noindent
{\bf Acknowledgements:} This work was supported by a grant from the Office of Basic Energy Sciences of 
the US Department of Energy under Contract No. FG05-86ER45234. Computations were made at the GATECH 
Center for Computational Materials Science.\\
~~~~~~\\
{\bf Author Contributions:} I.R. \& C.Y. performed the computations. C.Y., I.R. \& U.L. analyzed the 
results. C.Y. \& U.L. wrote the manuscript.\\
~~~~~~~~~~~~\\
{\bf Competing financial interests:} The authors declare no competing financial interests.\\
~~~~~~~~~~~~~\\
{\bf Correspondence} should be addressed to U.L.: (Uzi.Landman@physics.gatech.edu).


\end{document}